\documentstyle[12pt,aaspp4,epsfig]{article}
\def\spose#1{\hbox to 0pt{#1\hss}}
\newcommand\lsim{\mathrel{\spose{\lower 3pt\hbox{$\mathchar"218$}}
     \raise 2.0pt\hbox{$\mathchar"13C$}}}
\newcommand\gsim{\mathrel{\spose{\lower 3pt\hbox{$\mathchar"218$}}
     \raise 2.0pt\hbox{$\mathchar"13E$}}}

\def\ltsima{$\; \buildrel < \over \sim \;$}
\def\simlt{\lower.5ex\hbox{\ltsima}} % < over ~
\def\gtsima{$\; \buildrel > \over \sim \;$}
\def\simgt{\lower.5ex\hbox{\gtsima}} % > over ~

\def\V606{\hbox{$\rm V_{606}$}}
\def\I814{\hbox{$\rm I_{814}$}}
\def\I{\hbox{\rm I}}
\def\V{\hbox{$\rm V$}}

\newcommand{\etal} {{\it et~al.\ }}

\begin{document}

\title{HST Observations of the Host Galaxy of GRB970508}
 
\author{A. S. Fruchter$^1$,
E. Pian$^2$,
R. Gibbons$^1$, 
S. E. Thorsett$^{3,}$\altaffilmark{10},
H. Ferguson$^1$,
L. Petro$^1$,
K. C. Sahu$^1$,
M. Livio$^1$,
P. Caraveo$^4$,
F. Frontera$^2$,
C. Kouveliotou$^5$,
%J. Krist
D. Macchetto$^{1,6}$,
E. Palazzi$^2$,
H. Pedersen$^7$,
M. Tavani$^8$,
J. van Paradijs$^9$}

\affil{$^{1}$Space Telescope Science Institute, 3700 San Martin 
Drive, Baltimore, MD 21218, USA\\
$^{2}$Istituto di Tecnologie e Studio delle Radiazioni 
Extraterrestri, C.N.R., Via Gobetti 101, I-40129 Bologna, Italy\\
$^{3}$Joseph Henry Laboratories and Dept.\ of Physics, Princeton 
University, Princeton, NJ 08544, USA\\
$^{4}$Istituto di Fisica Cosmica e Tecnologie Relative,
C.N.R., Via Bassini 15, I-20133 Milano, Italy\\
$^5$NASA Marshall Space Flight Center, ES-84, Huntsville, AL 35812, USA\\
$^{6}$Affiliated to the Astrophysics Division, Space Science Department, 
European Space Agency\\
$^{7}$Copenhagen University Observatory, Juliane Maries Vej 30, D-2100, 
Copenhagen \"{A}, Denmark \\
$^{8}$Columbia Astrophysics Laboratory, Columbia 
University, New York, NY 10027, USA \\
$^{9}$Astronomical Institute ``Anton Pannekoek'', University
of Amsterdam, Kruislaan 403, 1098 SJ Amsterdam, The Netherlands}
%$^{4}$Department of Astronomy, Caltech, MS 105-24, Pasadena, CA 91125\\
%$^{6}$Dip. Fisica, Universit\`a di Ferrara, Via Paradiso 
%12, I-44100 Ferrara, Italy\\
%$^8$Space Telescope European Coordinating Facility, D-85748 Garching, 
%Germany\\}

\altaffiltext{10}{Alfred P. Sloan Research Fellow}
\begin{abstract}
We report on observations of the field of GRB~970508 made in early
August 1998, 454 days after outburst, with the STIS CCD camera
onboard the {\it Hubble Space Telescope.} The images, taken in open
filter (50CCD) mode, clearly reveal the presence of a galaxy which
was obscured in earlier (June 1997) HST images by emission from the
optical transient (OT).  The galaxy is regular in shape:
after correcting for the HST/STIS PSF, it is well-fitted 
by an exponential disk with a
scale length
of $0\farcs046 \pm 0\farcs006$ and an ellipticity of
$0.70 \pm 0.07$.  
All observations are marginally consistent with a continuous
decline in OT emission as $t^{-1.3}$ beginning two days after outburst;
however, we find no direct
evidence in the HST image for emission from the OT, and the surface
brightness profile of the galaxy is most regular if we assume that
the OT emission is negligible, suggesting that the OT may have
faded more rapidly at late times than is predicted
by the power-law decay.
Due to the wide bandwidth of the STIS clear mode, the estimated
magnitude of the galaxy is dependent on the galaxy spectrum that is
assumed. Using colors obtained from late-time ground-based
observations to constrain the spectrum, we find $\rm{V} = 25.4 \pm
0.15$, a few tenths of a magnitude brighter than earlier
ground-based estimates that were obtained by observing the total
light of the galaxy and the OT and then subtracting the estimated
OT brightness assuming it fades as a single power-law. 
This again suggests that the OT may have
faded faster at late time than the power-law predicts.  The position
of the OT agrees with that of the isophotal center of the galaxy to
$0\farcs01$ which, at the galaxy redshift $z = 0.83$, corresponds to
an offset from the center of the host of $\simlt 70$
pc. This remarkable agreement raises the possibility that the GRB
may have been associated with either an active galactic
nucleus or a nuclear starburst. 
\end{abstract}

\section*{Introduction}

The detection and rapid localization of GRB~970508 by the Gamma-Ray
Burst Monitor and the X-ray Wide Field Camera on {\it BeppoSAX}
(\cite{paa+98}) led to the identification of an optical counterpart
within four hours (\cite{bond97}) and, subsequently, to Keck
spectroscopy of the counterpart that revealed a system of absorption
lines at $z=0.835$ (\cite{mdk+97}). This lower limit on the GRB
redshift was the first direct constraint on the distance and energy
scale of a classical gamma-ray burst.  Because of its early discovery,
as well as the great interest attracted by the redshift measurement,
the fading counterpart of GRB~970508 has been more thoroughly studied
than any other GRB counterpart.  The optical light curve, for example,
has been intensively observed from a few hours to over
a year after the GRB. The optical flux reached a peak at R$\sim 19.8$
two days after the GRB, then began a power law decay, $t^{-\beta}$,
with $\beta=-1.141\pm0.014$, that continued for over one hundred days
(\cite{pfb+98a,ggv+98c}).  At that point, the decay curve began to
flatten (\cite{pfb+98b,pjg+98,bdkf98,zsb98b,szb+99}), as expected 
if the measured
flux were becoming dominated by light from a host galaxy.  GRB~970508
was also the first burst for which a radio counterpart was detected
(\cite{fkn+97,gwb+98b}).  The broadband (radio to X-ray) spectrum of
the afterglow (\cite{gwb+98}) provided strong support for the
synchrotron emitting shock model for afterglows (see, {\it e.g.},
Sari, Piran and Narayan 1998)\nocite{spn98}.

Despite the wealth of data on the GRB counterpart itself, the host
galaxy has proven a more difficult observational target.  Spectroscopy
has revealed [\ion{O}{2}] and [\ion{Ne}{3}] emission features, and
these, together with colors of the galaxy obtained by fitting
observations of the combined light from the OT and the galaxy, have
led to the suggestion that the host is an actively star-forming dwarf
galaxy (\cite{bdkf98,szb+99}). However, attempts to resolve the host
galaxy from the ground have proven fruitless.  Even early HST
observations, less than a month after outburst, found no evidence
for an extended source at the position of the optical transient, down
to faint levels,  $\rm{R} \gsim24.5$ (\cite{pfb+98a}).  In this Letter, we
describe HST observations taken more than a year after outburst, which
have finally allowed us to resolve the host galaxy of GRB~970508.
These show that GRB~970508 occurred remarkably close---within about
70\,pc---of the host galaxy center, and suggest that the brightness of
the OT may have fallen faster at late times than would be predicted by
a simple power-law fit.  Finally, we discuss the implications of these
observations for understanding the progenitor objects and energetics
of GRBs.

\section*{Observations, Image Analysis, and Results}

%\subsection*{The STIS Images}

The field of GRB~970508 was imaged during four HST orbits in 1998
August 5.78--6.03\,UT, using the STIS CCD in Clear Aperture (50CCD)
mode.  Two exposures of 1446\,s each were taken at each of four dither
positions for a total exposure time of 11,568\,s.  The images were
bias and dark subtracted, and flat-fielded using the STIS pipeline.
The final image was created and cleaned of cosmic rays and hot pixels
using the variable pixel linear reconstruction algorithm (a.k.a.\
Drizzle) developed for the Hubble Deep Field (\cite{wms96,fh97}).  An
output pixel size of $0\farcs025$ across (one-half the size of the
detector pixels on the sky) and a ``pixfrac" of 0.6 were used.  The
(small) geometric distortion of the STIS CCD (\cite{mb97}) was removed
during the drizzling process.  A section of the final image is shown
in Fig.~1.
%In order to make the faint extended nebula more easily visible,
%we have also block-averaged the high resolution image onto pixels
%$0\farcs1$ across.  Both  of these images are shown in
%Figure 1.

The total emission from the OT and galaxy were measured by summing the
counts in a box $1\farcs5$ on a side and subtracting the local sky.
We find $3.13 \pm 0.12$ counts per second in the aperture.  The
photometric calibration of the images was performed using the
synthetic photometry package SYNPHOT in IRAF/STSDAS;
however, a 12\%  aperture correction has been
applied \cite{lan97b} to account for light lost to
large-angle scattering.  The STIS CCD in clear aperture mode has a broad
bandpass, with a significant response from 200 to 900\,nm that peaks
near 600\,nm.  As a result, STIS instrumental magnitudes are best
translated into the standard filter set by quoting the
result as a V magnitude; however, knowledge of an object's
intrinsic spectrum is required for an accurate conversion to the
standard filter system.  Using a spectral energy distribution (SED)
flat in $f(\nu)$ one finds $\rm{V} = 25.1 \pm 0.1$.
But, as mentioned in the introduction, ground-based observers have
fitted for the host galaxy magnitude under the assumption that the
power-law index of decay of the OT has been constant with time.  We
can therefore estimate the V magnitude using the color information
from these observations.  Although the estimated galactic magnitudes
have changed with time (a point we will return to later), all
observers have found a blue host, and Sokolov \etal suggest that the
galaxy colors are best fit by an object intermediate
between an Scd and an irregular (Im) redshifted to $z=0.83$.  Using
either the measured galaxy colors, obtained by a rough averaging of
the values obtained by previous observers
(\cite{bdkf98,zsb98b,szb+99}), or an SED created by interpolating
between Coleman, Weedman and Wu (1980) \nocite{cww80} Scd and Im SEDs
and redshifting to $z=0.83$, we estimate $\rm{V} = 25.40 \pm 0.15$,
where the error is dominated by our uncertainty over the SED.  This,
however, represents the sum of the emission from the host galaxy and
any remnant of the OT.  We next place a limit on the magnitude of the
OT.
%We can reproduce the $\rm{V} \sim 25.8$
%found by the two groups if we assume the SED is that of the Coleman,
%Weedman and Wu (1980) \nocite{cww80} Sbc at redshifted of $z=0.8$.
%However, the restframe UV colors of the host implied by the
%ground-based observations are considerably bluer than that of this SED
%(though the magnitude of the 4000~\AA\ break is comparable).  A
%discrepancy could arise between our observations and those from the
%ground, if the OT were to deviate from the power-law assumed in the
%ground-based fits.  Therefore, we now turn to measuring the
%contribution of the OT to the total magnitude.

In order to register the position of the OT on the late-time image,
 the
positions of nine compact sources were found on both the June 1997
(\cite{pfb+98a}) and July 1998 drizzled images.  A shift (in $x$ and
$y$) and rotation were then fit between the two images using the IRAF
task {\bf geomap}.  The accuracy of this transformation was checked by
comparing the observed and predicted positions of four bright,
point-like sources.  An r.m.s.\ scatter of $0.25$ drizzled pixels
($0\farcs006$) was found in each coordinate, for a position
uncertainty $< 0\farcs01$.
When the position of the OT on the June 1997 image was transformed
to that of the July 1998 image using the shift and rotation measured,
we found it to be exactly at the center of the host.  To verify this
observation, we fitted the host galaxy with elliptical isophotes using
the IRAF task {\bf ellipse}.  We find that the isophotal center of the
galaxy is stable as a function of radius and agrees with the predicted
position of the OT to better than our astrometric error of
$0\farcs01$.

In Fig.~2 we show a plot of the measured surface brightness profile of
the galaxy compared with an $r^{1/4}$ model and an exponential
disk model.  In both cases, we have convolved the model with the STIS
PSF. In addition to the measured surface brightness profile, we show
that profile after subtracting an estimated remnant OT.  To do this,
we went back to the June 1997 observation and scaled and subtracted a
STIS PSF until the remaining counts in a circle of radius four
drizzled pixels equaled that in the same region of the late-time
image.  This PSF was then used as the estimate of the OT at 24.7 days
after outburst,  was then itself scaled using the $t^{-1.14}$
power-law found in Pian \etal (1998)\nocite{pfb+98a} to the late time,
454 days after outburst.  When subtracted from the galaxy, this
estimate of the OT produced a clear ``hole'' in the center of the
host.  Under the assumption that galaxies (convolved to $\sim 500$ pc
resolution by the PSF) should have surface brightness profiles rising
toward the center, we reject this subtraction.  The largest
subtraction consistent with a roughly continuously rising surface
brightness profile is shown in Fig.~2.  Therefore we have subtracted a PSF
scaled as $t^{-1.3}$ between the two HST observations.  This power-law
is $2 \sigma$ below the power-law reported by Pian \etal (1998), but
agrees well with that found by Bloom \etal (1998) and is within $2
\sigma$ of that found by Zharikov \etal (1998). (We note the Pian
\etal fit was slightly contaminated by the then-unmeasured light from
the host galaxy).

As can be seen from Figure~2, the surface brightness profile of the
host galaxy is a far better fit to by exponential disk model than by an
$r^{1/4}$ law.  The best fit exponential model shown has a scale
length $= 0\farcs046 \pm 0\farcs006$ and ellipticity $= 0.70 \pm 0.07$.  
It has then been convolved with an estimate of the STIS PSF,
produced using the HST Tiny Tim software (\cite{khb92}) (results
obtained when the image is convolved using a stellar PSF are quite
similar).  This convolution produces a model which can be approximated
by an exponential disk with scale length $\sim 0\farcs060$ and
ellipticity $\sim 0.3$.  A true exponential disk plotted as magnitude
versus radius would, of course, have a surface brightness profile that is
a straight line; however, at its core, the surface brightness of the
observed galaxy is averaged over the width of the PSF, and at large
radii the true light of the galaxy is overwhelmed by light scattered
from the center.  It is worth noting that given the large eccentricity
observed, the poor fit of the $r^{1/4}$ law is not unexpected.  In
spite of their names, ellipticals rarely have ellipticities
approaching $0.7$.

The fit between the galaxy models and the data is substantially better
when {\it no} OT is subtracted, than when we remove an OT scaled as
$t^{-1.3}$.  Nonetheless, as can be seen in Figure~3, this power-law
largely fits the available ground-based R-band photometry.  For this
figure, a galaxy magnitude of $\rm{R} = 25.2$ has been removed from
previous photometry.  This corresponds to a flat (in $f(\nu)$),
{\i.e.} very blue, galaxy spectrum between R and V.  The colors found
by Sokolov \etal are somewhat redder; however, their V galaxy
magnitude $25.16 \pm 0.16$ is somewhat fainter than ours.  In July
1998, an OT falling as $t^{-1.3}$ would have $\rm{V} \sim 25.8$,
implying a corrected galactic magnitude of $\rm{V} = 25.5 \pm0.15$.
However, there has been a continuing trend among the ground-based
estimates of the host magnitude.  The later the data used to fit the
host galaxy, the fainter the host was found to be
(\cite{bdkf98,zsb98b,szb+99}).  The differences are visible in all
bands (B, V, R and I), and are typically at the $2 \sigma$ level.
Furthermore, the preference of our surface brightness fit for no
continuing emission from the OT, and the prevalence of upper limits,
rather than detections beyond day 150 in Figure~3, all suggest a single
conclusion---the OT may have faded much more rapidly than $t^{-1.3}$
after day $\sim 100$.

\section*{Discussion}

Our imaging has revealed the faint galaxy host of GRB~970508.  We find
that the OT is located, within astrometric errors of order
$0\farcs01$, at the isophotal center of the host.  At the redshift of
GRB~970508, $z=0.83$, this corresponds to an offset from the nucleus
of $\simlt 70$~pc.  The surface brightness profile of the host
better fits an exponential disk than the $r^{1/4}$ profile of an
elliptical, and agrees best when no OT is assumed to be adding to the
profile; however, we cannot rule out a power-law decay of the OT as
$t^{-\beta}$, where $\beta \geq 1.3$.  Both our data, and the
ground-based observations, tend to support a steepening of the early
power-law decay curve sometime after day $\sim 100$.  Such a break is
naturally expected when the expanding fireball has swept up the
material from the ISM comparable in rest mass to the energy of the
initial explosion at:
\begin{displaymath}
t\approx 1\mbox{\,yr}\left(\frac{E_{52}}{n}\right)^{1/3},
\end{displaymath}
where $E_{52}$ is the initial energy of the explosion in units of
$10^{52}$~erg and $n$ is the density of the surrounding medium in
protons per cm$^3$ (\cite{wrm97}).     Wijers and Galama (1999)
\nocite{wg99} have used the multi-wavelength observations of
the afterglow emission of GRB~970508 to estimate the physical
parameters of the burst and its surrounding interstellar medium,
assuming that the afterglow is dominated by synchrotron emission.
They find a total burst energy of $\sim 4 \times 10^{52}$ and
$n \sim 0.04$.  These values would cause us to expect a break
somewhat after one year. However, the uncertainties in these estimated
parameters are large (perhaps an order of magnitude,
Wijers, private communication). Furthermore the above calculation
does not take into account the significant time-dilation in the
early part of the expansion, thus overestimating the time till
the break.  Therefore, we believe that all observations are
consistent with the possible break in the light curve between
100 and 200 days after outburst.
%(As noted by Panaitescu and M\'esz\'aros
%(1998),\nocite{pm98} this break can be mitigated by a significantly
%non-spherical outflow geometry.)

Although the precise behavior with time of the OT is uncertain, 
its position on the host is not.  The extraordinary coincidence of the
OT with the isophotal center of the host galaxy raises the question of
whether the GRB is related to the galactic nucleus, either through a
nuclear starburst or an AGN.  The Keck spectroscopy by Bloom \etal
(1998) shows strong [\ion{O}{2}] and [\ion{Ne}{3}], both of which are
present in galaxies with active nuclei. However, in a large
spectroscopic sample of galaxies (\cite{mck95,skc95}), no ellipticals
or spirals without AGN show [\ion{Ne}{3}].  About one-third of the
starbursts in the sample show this line, and these are by and large
the most active starbursts in the group.  Furthermore, only the most
extreme starbursts, and the Seyferts, have a [\ion{Ne}{3}] equivalent
width or the large [\ion{Ne}{3}] to [\ion{O}{2}] ratio (indicative of
a temperature in excess of 40,000 K) seen in this galaxy.  Thus, the
spectroscopic evidence does not allow us to distinguish between a host
which possesses an AGN and one which is simply showing signs of
vigorous star formation.  Nonetheless, we tend to prefer the latter
explanation for two reasons.  
First, if cosmological GRBs are produced by a single mechanism, that
mechanism is unrelated to AGN.  The OT of GRB~970228 is located
at the very edge of a galactic disk (\cite{fpt+99}).   Furthermore,
HST images of other GRBs (\cite{odk+98,bfk+98,ftm+99}), while
less conclusive, all tend to discourage an AGN interpretation.
Secondly, our recent work has shown that other GRB hosts possess
unusually blue optical-to-infrared colors, implying that these
galaxies are actively star-forming (Fruchter \etal 1999b).  
NICMOS imaging should soon allow us to determine whether this
is also the case for the host galaxy of GRB~970508.  Until then, we
note that in many ways this host galaxy has a strong resemblance to
the classic nearby starburst dwarf NGC~5253, in its integrated colors,
morphology and line-strengths (\cite{mck95,skc95}).  Furthermore,
NGC~5253 has a hot, young star cluster in its nucleus, suggesting that
the resemblance may be very good indeed, by providing a natural
explanation for the location of the OT.

The hosts of four GRBs (970228, 970508, 971214 and 990123) have now
been imaged and clearly resolved by HST
(\cite{slp+97,fpt+99,odk+98,ftm+99,bod+99}). 
%and the host of GRB~980703 has
%been resolved from the ground (\cite{bfk+98}).  Not true, it seems.
In each case, the OT
is superposed on the stellar field.  We note that this may be a result
of selection effects and not the true distribution of GRBs with
respect to host galaxies, since all of these GRBs were localized by
detection of an OT, which itself may require the presence of a dense
external working surface such as an ISM (\cite{pr93,mr97a}).
Furthermore, only $\sim 50 \%$ of the GRB localizations by the
Beppo-Sax satellite have resulted in the discovery of an optical
transient.  This fraction is consistent with a model of GRB
formation from the merger of neutron-star-neutron-star binaries, since
a substantial fraction of neutron-star-neutron-star binaries are likely to be
ejected from the galaxy by the momentum imparted to the neutron-stars
at birth (\cite{bsp98,lsp+98}).  However, it is not immediately clear that
star-formation can properly account for the fraction of GRBs
detected in the optical.  Local estimates of
dust obscuration in star-forming galaxies (\cite{ch99}), as well as
some estimates of the same effect in high-redshift galaxies
(\cite{pks+98,bsik99}), suggest that about one-third of the light
emitted in the UV escapes from starforming galaxies before being
absorbed by dust and reprocessed to IR or radio wavelengths.  (We
typically view the OTs of GRBs in the UV rest wavelength since they
have observed redshifts between 0.8 and 3.4, see also Hogg and
Fruchter 1999.)  A reduction by a
factor of three of the light emitted by GRBs would be roughly
consistent
with what we observe --- about one-half of GRBs are missing,
and of order one-half of those observed have redder spectra than
expected based on the afterglow theory (\cite{bfk+98,fpt+99,hthc98}),
perhaps suggesting the presence of moderate extinction.  However,
other authors (\cite{mhl+97}) have claimed significantly higher
absorption by dust at high redshift.  And deep sub-millimeter
observations of several high Galactic latitude fields
(\cite{hsd+98,bcs+98}) have suggested that a few obscured objects in
each field which are undetectable in the optical could be producing
more stars than all of the galaxies visible in the optical.
If these more extreme estimates of the importance of dust
obscuration are correct, and GRBs are related to star formation, it
may be difficult to explain the success optical observers have had in
finding OTs---especially ones like 970508, which is quite probably at
the nucleus of a highly inclined starburst galaxy, yet whose color in
the rest-frame UV (\cite{pfb+98a}) shows no sign of significant
extinction.

\acknowledgements

We thank Bob Williams for allocating Director's Discretionary time to
observe GRB~970508 using STIS and John Krist for preparing Tiny
Tim STIS PSFs for us.

\section*{Figures}

\begin{figure}
\centerline{\psfig{file=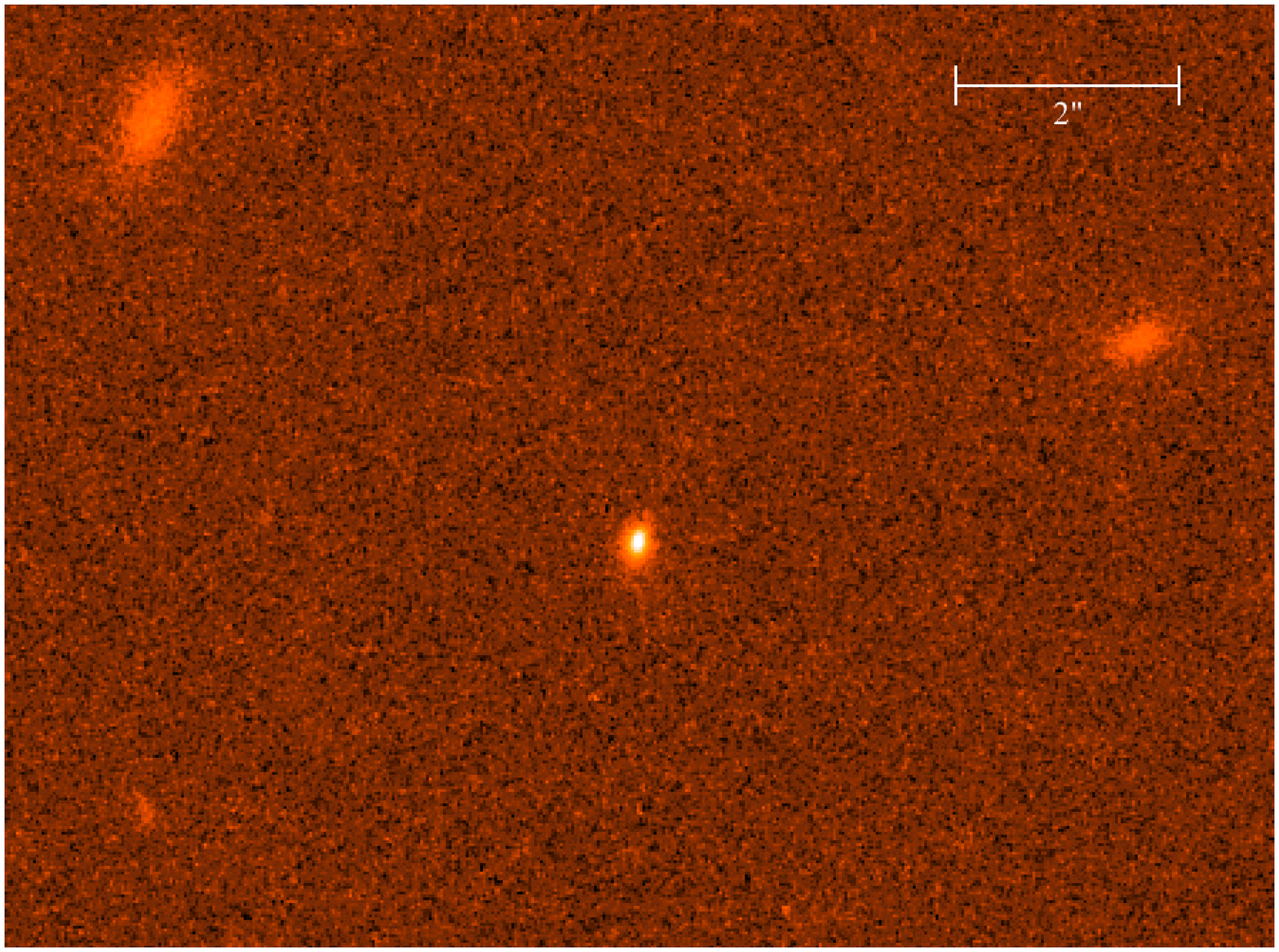,width=5.5in,angle=-90}}
\vspace{10pt}
\caption{ A section of the STIS CCD image of the field around GRB~970508. 
North is up, East is to the left.  The host galaxy is near the center of
the field.  The estimated position of the OT agrees with the center of the
host to $0\farcs01$.}

\end{figure}

\begin{figure}
\centerline{\psfig{file=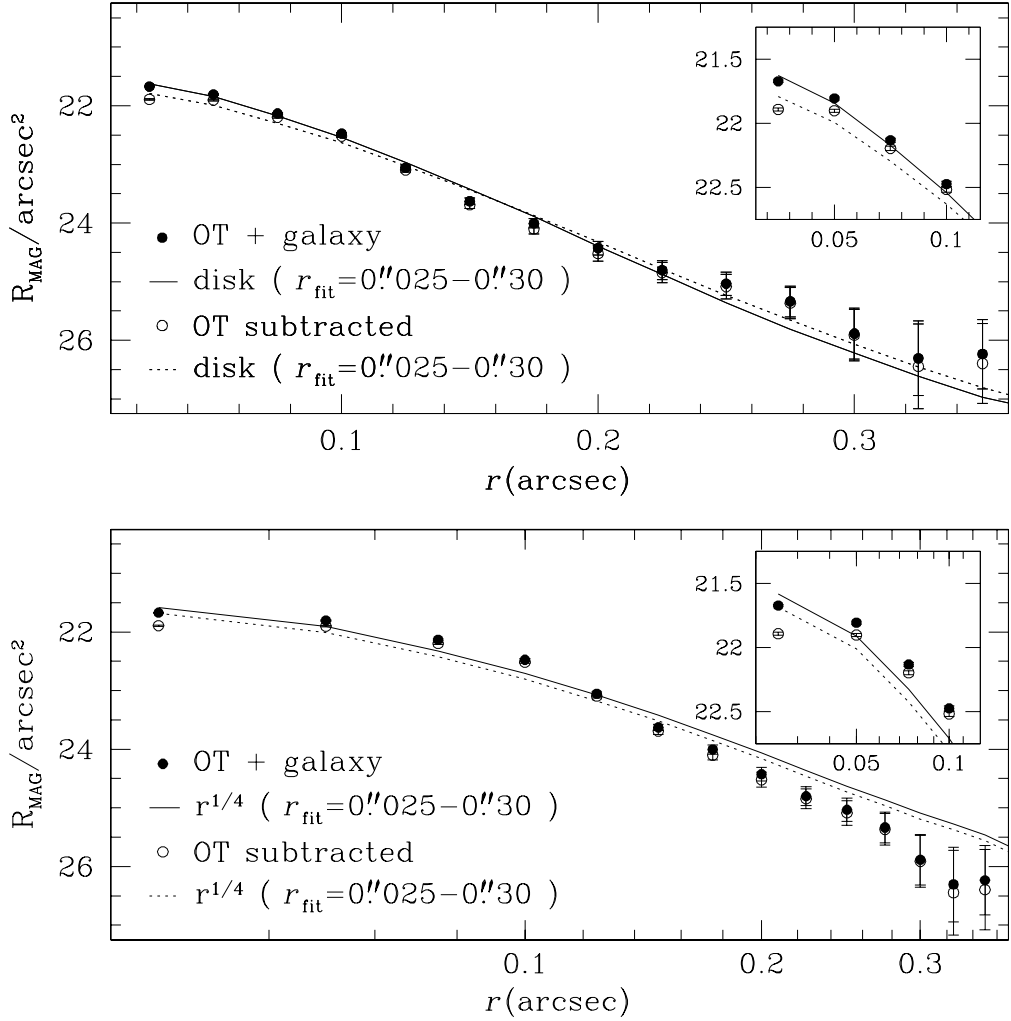,width=5.5in}}
\vspace{10pt}
\caption{The surface brightness profile of the host galaxy with and without
the OT removed.  The upper plot shows the measured isophotes of the host
compared with our best-fitting exponential-disk model.  The
solid points are the measured values on the original image.  
The open points show the values measured after subtraction of an OT
with $V=27.8$, corresponding to a decay of the OT with time
as $t^{-1.3}$.  The model has been fit to both the observed data (solid
line) and the galaxy with OT subtracted (dashed lines).  The fit was
done over radii from 1 to 12 drizzled pixels ($0\farcs025$ to $0\farcs30$).
The lower graph shows the same information for an $r^{1/4}$-law fit.}
\end{figure}

\begin{figure}
\centerline{\psfig{file=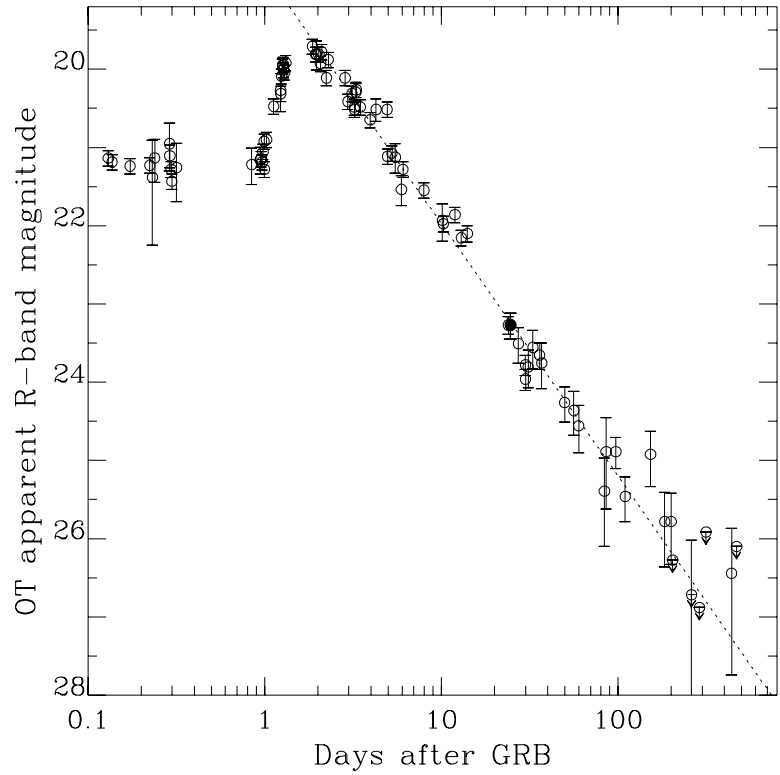,width=5.5in}}
\vspace{10pt}
\caption{The R-band light curve of the OT of GRB~970508.  A constant
host R magnitude of 25.2 has been subtracted from all observed
values.  The solid line is a power-law with slope $t^{-1.3}$.  Two-sigma
upper-limits are marked by a downward-pointing arrow.  The data have
been taken from Pian \etal (1997), Bloom \etal (1998), Zharikov \etal
(1998) and  Sokolov \etal (1999) and references therein.}
\end{figure}
\end{document}